\documentstyle{article}

\begin{document}

\title{Inclusive and Exclusive Semi-Leptonic Decays\\
of $B$ Mesons}
\author{M. Phipps \\ Physics Department,
McGill University, Montreal}
\maketitle

\begin{abstract}
In this paper, the semileptonic decays of heavy mesons are treated
fully relativistically.
By means of an effective vertex, the effect of Fermi momentum
are included both at the inclusive and at the exclusive levels,
and the spin of both parent and daughter particles are taken into
account.  The differential decay rates
with respect to the lepton energy and momentum transfer are compared
with data from ARGUS and CLEO.
\end{abstract}

\section*{Introduction}

There are several reasons why semileptonic $B$ decays are of interest.
For one thing, the are a small variety of decay products, namely
those which contain charm quarks ($D$, $D^*$, $D^{**}$, etc.)
and those which do not ($\pi$'s etc.).  $V_{ub}/V_{cb}$ can be determined
from the relative number of these decays.  In addition,
the heavy masses of the b and c quarks suggest that one
might be able to apply perturbative QCD to calculate the strong
corrections to these processes. This hope has been recently
formalised in Heavy Quark Effective Theory\cite{HQET}.  Finally,
there's the fact that theoretical uncertainties are much smaller than in
non-leptonic decays which contain a wider variety of hadronic decay
products.

If we write out a parameterisation for the CKM Matrix, we see that it
depends on a complex phase which is resposible for CP violation
in the standard model.  The magnitudes of the values for the CKM matrix
elements place limits on the size of this phase and, thus, on the amount
of CP violation in the standard model.
According to the particle data group,
$V_{ub}=.0035\pm.0015$ and $V_{cb}=.040\pm.08$.  Recent
values of $V_{ub}$ and $V_{cb}$ in the literature fall in this range
[3-8].

In studying semi-leptonic decays, the first approximation is to neglect
QCD and use a spectator model in which the up quark of the $B$ meson is
not involved in the decay except to recombine with the charm quark.  In
such a model, the decay of the $B$ meson into a $D$ meson reduces to that
of the decay of a bottom quark into a charm quark.  Quantities that can be
determined directly from experimental data include the square of the
momentum transfer,
\begin{eqnarray}
Q^2&=&(B-D)^2\nonumber\\
&=&m_B^2+m_D^2-2E_BE_D+2\vec p_B\cdot\vec p_D,
\end{eqnarray}
and the lepton energy, $E_l$.

\section*{Inclusive Case}

Now that we have a process involving quanties that can be determined
from experiment, we would want to come up with a theory that relates
these quanties.  One such model was the one devised by Altarelli, Cabbibo,
Corbo, Maiani and Martinelli in 1982\cite{Alt}.  In this model, the
bottom quark was assumed to be on shell and, thus, given by
\begin{eqnarray}
m_b^2&=&(B-u)^2\nonumber\\
&=&m_B^2+m_u^2-2m_B\sqrt{p_u^2+m_u^2}
\end{eqnarray}
in the $B$ rest frame.  This assumption has the advantage that it
avoids having the decay rate depend on an arbitrary overall $1/m_b^5$
that appears in a purely partonic treatment of these decays\cite{Bar}
The up quark momentum was then assumed to
obey a Gaussian distribution,
\begin{equation}
\phi(p)={1\over\pi^{3\over2}p_f^3}exp\left({-p^2\over p_f^2}\right)
\end{equation}
which is normalised according to
\begin{equation}
\int d^3p \phi(p)=1
\end{equation}
and is thus fixed up to an adjustable parameter, $p_f$, known as
the Fermi momentum.

In our case, we want to consider the up quark as being more than a mere
spectator: instead we write the effective
$\overline uBb$ vertex as $\gamma_5 V_B(p_u)$.
The decay rate in this model is
\begin{equation}
\Gamma(B\to\overline ucl\overline\nu)={N|V_{cb}|^2|V_B|^2\over2m_B(2\pi)^8}\int
{d^3p_c\over 2E_c}{d^3p_l\over 2E_l}
{d^3p_\nu\over 2E_\nu}{d^3p_u\over2E_u}\phi(p_u)|M|^2
\delta^4(B-u-c-l-\nu)
\end{equation}
where
\begin{equation}
|M|^2=G_F^2L_{\alpha\beta}H^{\alpha\beta}
\end{equation}
and
\begin{equation}
H^{\alpha\beta}=Tr[\gamma^\alpha(1-\gamma_5)(c\kern -5.5pt /+m_c)\gamma^\beta
(1-\gamma_5)(b\kern -5.5pt /+m_b)\gamma_5(-u\kern -5.5pt /+m_u)\gamma_5
(b\kern -5.5pt /+m_b)].
\end{equation}

It turns out this this integral is difficult to evaluate, the problem being
that the up and charm quarks are not being assumed to combine into a specific
meson.  Instead, the quarks are combining to form a cluster $X$ with four
momentum $X=u+c$ and mass $m_X^2=(u+c)^2$.  This $m_X$ is arbitrary save for
the fact that, experimentally, $m_X>m_D=1.8963$ GeV while energy conservation
requires that $m_X<m_B$.  As a result, we will want to rewrite the hadronic
phase space so that $m_X$ is integrated over this range.
Using standard cluster decomposition techniques,
the decay rate becomes:
\begin{eqnarray*}
\Gamma(B\to X l\overline\nu)={N|V_{cb}|^2|V_B|^2\over2m_B}\int&|M|^2&\phi(p_u)
{1\over(2\pi)^2}{d^4p_c}{d^4p_u}\delta^4(X-u-c)\delta(c^2-m_c^2)
\delta(u^2-m_u^2)\nonumber\\
&\times&{1\over(2\pi)^2}{d^4p_l}{d^4p_\nu}\delta^4(Q-l-\nu)\delta(l^2)
\delta(\nu^2)\nonumber\\
&\times&{1\over(2\pi)^2}{d^4Q}{d^4X}\delta^4(B-Q-X)\delta(Q\cdot Q-Q^2)
\delta(X^2-m_X^2)\nonumber\\
&\times&{1\over(2\pi)^2}dQ^2dm_X^2
\end{eqnarray*}

Using\cite{Bar}
\begin{eqnarray}
\int d_2(X\to ab)&=&(\pi/2)\lambda^{1\over2}(1,a^2/X^2,
b^2/X^2){d\Omega\over 2\pi},
\end{eqnarray}
we get
\begin{equation}
{1\over(2\pi)^2}{d^4Q}{d^4X}\delta^4(B-Q-X)\delta(Q\cdot Q-Q^2)
\delta(X^2-m_X^2)={1\over2\pi}{p_Q\over 2m_B}
\end{equation}
where $p_Q=\lambda^{1\over2}(1,X^2/m_B^2,Q^2/m_B^2)m_B=p_X$.
The remaining delta functions are
\begin{eqnarray*}
\delta(c^2-m_c^2)&=&\delta((X-u)^2-m_c^2)\\
&=&\delta(X^2+m_u^2-2E_XE_u+2p_Xp_ucos\theta_{Xu}-m_c^2)
\end{eqnarray*}
and
\begin{eqnarray*}
\delta(\nu^2)&=&\delta((Q-l)^2)\\
&=&\delta(Q^2-2E_QE_l+2p_QE_lcos\theta_{Ql})
\end{eqnarray*}
These cancel with the cosine integrations in $d^3p_u$ and $d^3p_l$.
The final expression for the decay rate is, thus,
\begin{equation}
\Gamma(B\to X l\overline\nu)={N|V_{cb}|^2
V_B^2\over(2\pi)^6(2m_B)^2}\int|M|^2\phi(p_u)
{p_udp_udE_ld\phi\over16p_QE_u}dQ^2dm_X^2
\end{equation}

If we now compare this formula with data from ARGUS\cite{A1}\cite{A2} and
CLEO\cite{C1} then we find, after minimising with
respect to the parameters $m_u$, $m_c$, $m_b$, $p_f$ and $|V_{ub}|/|V_{cb}|$,
that we get a good fit for parameters in the ranges
\begin{eqnarray*}
m_u&=&.13\pm.38\ {\rm GeV}\\
m_c&=&1.4\pm.4\ {\rm GeV}\\
m_b&=&4.9\pm.3\ {\rm GeV}\\
p_f&=&.5\pm.1\ {\rm GeV}\\
{\rm and}\ V_{ub}/V_{cb}&=&.07\pm.05
\end{eqnarray*}

Note that the ARGUS and CLEO data include
contributions from $b\to c l\overline\nu$ and $b\to u l\overline\nu$
decays.  In each case,
the measured electrons were separated into different categories
including electrons from non-$\Upsilon(4S)$ events,
$\psi$ or $\psi(2S)$ decay,
$\tau$ decay or semileptonic $D_s$ decay,
semileptonic $D$ decay,
$\pi^0\to e^+e^-$ decay and
semileptonic $B$ decay,
the latter being the ones that are used to make these plots.
Additional background comes from having hadrons misidentified as electrons.
Note that for $E_l > 2.4$ GeV, electrons from $B$ decay can only come
from charmless semi-leptonic decays.

Using these parameters and taking the areas under
the curves gives us the branching ratio $Br(B\to c
\overline u l\overline\nu)$=10.09\% and $Br(B\to u\overline u l\overline\nu)$
=.16\%.  Using
\begin{equation}
\Gamma(B\to\overline ucl\overline\nu)=Br(B\to\overline ucl\overline\nu)/
\Gamma_B
\end{equation}
and knowing\cite{PDG} that $\Gamma_B=(1.52\pm.11)\times10^{-12}(1.52
\times10^{24})$ GeV, $|V_{cb}|$ can be calculated to be $.034\pm.003$.

\section*{Exclusive Case}

In the spectator model, one can differentiate between $D$ and $D^*$
mesons according to whether the daughter meson has spin 0 or 1.
That one can do this was overlooked in a recent paper by V. Barger {\em
et al} that attempted to differentiate between different decay products
in the differential $m_X$ distribution\cite{Kim}.
Mahiko Suzuki\cite{Suz} used this observation to calculate exclusive
rates at zero Fermi momentum.  In this frame,
\begin{equation}
H^{\alpha\beta}=M_0^{\alpha}M_0^{\beta}+M_1^{\alpha}M_1^{\beta}
\end{equation}
where
\begin{eqnarray}
M_0^\lambda&\propto&Tr\left[(c\kern-4.5 pt / +
m)\gamma^\lambda
(1-\gamma_5)(b\kern-5. pt / + M)\right]
/[4M\{2m(E_c+m)\}^{1\over2}]
\end{eqnarray}
and
\begin{eqnarray}
M_1^\lambda&\propto&Tr\left[(c\kern-4.5 pt / +
m)\gamma_5\epsilon\kern-5. pt /\gamma^\lambda
(1-\gamma_5)(b\kern-5. pt / + M)
\right]/[4M\{2m(E_c+m)\}^{1\over2}].
\end{eqnarray}
Here $\epsilon_\lambda$ represents the three polarisations
satisfying $\epsilon_\lambda c^\lambda=0$.  In the rest frame of $c$,
$\epsilon_\lambda$ is, therefore, given by
\begin{eqnarray}
&\epsilon_\lambda^{(T)}=(0,1,0,0),&(0,0,1,0) \nonumber \\
&\epsilon_\lambda^{(L)}=(0,0,0,1).&
\end{eqnarray}
where $(T)$ and $(L)$ signify transverse and longitudinal polarisations,
repectively.

In the case where the $b$ is not at rest in the $B$ rest frame
$\epsilon$ is defined specifically in the $B$ rest frame.
Now, if instead of considering the light quark as a spectator, we
treat it
as an intermediate decay product in an effective theory involving a
$\overline ubc$ loop, then the relevant traces are
\begin{eqnarray}
&M_0^\lambda\propto Tr\left[(c\kern-4.5 pt / +
m_c)\gamma^\lambda
(1-\gamma_5)(b\kern-5. pt / + m_b)\gamma_5(-u\kern-5.5 pt / +
m_s)\gamma_5
\right]& \\
&M_1^\lambda\propto Tr\left[(c\kern-4.5 pt / +
m_c)\gamma^\lambda
(1-\gamma_5)(b\kern-5. pt / + m_b)\gamma_5(-u\kern-5.5 pt / +
m_s)\epsilon
\kern-4.5 pt /\right]&
\end{eqnarray}
The Suzuki matrix elements are reproduced
as  $\vec p$ goes to zero.

Starting with the $B$ meson at rest,
\begin{equation}
\Gamma(B\to D l\overline\nu)={1\over{2m_B(2\pi)^5}}\int {d^3p_D\over
2E_D}{d^3p_l\over
2E_l}
{d^3p_\nu\over 2E_\nu}|S|^2\delta^4(B-D-l-\nu)
\end{equation}
where
\begin{equation}
S= {N^{1\over2}G_FV_{cb}V_BV_D\over2\pi}\int{d^3p_b\over2E_u}|\phi^*(p_u)
\psi(t_u)|^{1\over2}M
\end{equation}
and where $\vec p_u(\vec p_u^\prime)$ and $\vec t_u(\vec t_u^\prime)$
are the up quark momenta in the $B$ and $D$ rest frames, respectively,
$V_B$ and $V_D$ are the vertex constants and $N$
is a normalisation.  The wavefunctions $\phi(p_u)$ and $\psi(t_u)$ are
\begin{eqnarray}
\phi(p_u)={1\over\pi^{3\over2}p_f^3}exp\left({-p_u^2\over p_f^2}\right)&
{\rm and}&\psi(t_u)={1\over\pi^{3\over2}t_f^3}exp\left({-t_u^2\over t_f^2}
\right)
\end{eqnarray}
where $p_f$ and $t_f$ are independent adjustable parameters.  $t_u$ is
given by
\begin{eqnarray}
\vec t_u^2&=&E_t^2-m_u^2\nonumber\\
&=&[(E_uE_D-\vec p_u\cdot\vec p_D)/m_D]^2-m_u^2
\end{eqnarray}
where $E_t$ is the energy of the up quark in the $D$ rest frame.

The phase space simplifies as follows:
\begin{eqnarray}
&d_3(B\to D l\overline\nu)\propto{d^3p_D\over 2E_D}{d^3p_l\over 2E_l}
{d^3p_\nu\over 2E_\nu}{d^3p_u\over 2E_u}{d^3p_u^\prime\over 2E_u^\prime}
\delta^4(B-D-l-\nu)&\\
&={\pi\over8}{dp_Dp_D^2dp_lp_l^2\over E_DE_lE_uE_u^\prime}dcos\theta_l
d\phi_l\delta(\nu^2)dp_up_u^2dcos\theta_ud\phi_udp_u^\prime p_u^{\prime2}
dcos\theta_u^\prime d\phi_u^\prime.&
\end{eqnarray}
$B-D-l-\nu=0$ and $\nu^2=0$ implies
\begin{eqnarray}
&(B-D-l)^2=0&\\
&\Rightarrow 2p_DE_lcos\phi_l=2E_DE_l-2m_B(E_D+E_l)+(m_D^2+m_B^2)&
\end{eqnarray}
so the phase space becomes
\begin{equation}
d_s(B\to D l\overline\nu)\propto{\pi^2p_D\over8E_D}{dp_DdE_l\over E_u
E_u^\prime}dp_up_u^2dcos\theta_ud\psi_udp_u^\prime p_u^{\prime2}
dcos\theta_u^\prime d\psi_u^\prime.
\end{equation}

If we now insert into this model the parameters given in the previous section
we run into problems: it turns out that we only get agreement with ARGUS
\cite{A3} data for low values of $Q^2$.
This is presumably due to final-state interactions, which in a perturbative
QCD framework are expected to grow as one approaches the end-point of the $Q^2$
distribution.
In the exclusive case, QCD corrections are restricted to those which do
not create additional hadrons, that is quark propagator self-corrections
and vertex corrections.  Corrections to the $cD\overline u$ vertex are of
particular interest because they provide a phenomenological explanation
for the discrepancy: the exchange of a gluon between the up and charm quark
can reduce their relative momentum, allowing them to combine to form a
$D$ or $D^*$ meson.

\section*{Conclusion}

This model effectively describes the dependence of
both inclusive and exclusive semileptonic $B$ decays on the Fermi
momentum of the constituent quarks.  The parameters that arise naturally
in this model agree with those used in other models.

Given that this model describes both inclusive and exclusive decays,
we can estimate the rate of semileptonic $B$ decays into $D^{**}$
mesons or clusters consisting of $D$'s or $D^{*}$ and $\pi$'s
by subtracting the exclusive rates into $D$ and $D^{*}$ from the
inclusive semileptonic $B$ decays into charmed mesons.  Experimentally,
this rate is found to be between 33\% and 41\% of the total semileptonic
rate\cite{who1}\cite{who2}.  This model would appear to have
the best chance of accounting for
all possible semileptonic decay products of $B$ mesons.

%Figures

\bigskip

Figure 1: ${dBr\over dE_l}$ for $m_u$=.13 GeV, $m_c$=1.4 GeV, $m_b$=4.9
GeV, $p_f$=.5 GeV and $V_{ub}/V_{cb}\approx$.07 with data from
ARGUS[11][12] and CLEO[13]

\bigskip

Figure 2: ${dBr\over Q^2}$ for $m_u$=.13 GeV, $m_c$=1.4 GeV, $m_b$=4.9 GeV,
$p_f$=.5 GeV and $V_{ub}/V_{cb}\approx$.07 with data from ARGUS[17]


\begin{thebibliography}{99}
\bibitem{PDG} Particle Data Group, {\em Phys. Rev.} {\bf D50} (1994), 1173.
\bibitem{HQET} M. Neubert, CERN Preprint \#7396 (1994).
\bibitem{Li} H.-N. Li, and H.-L. Yu, Chung-Cheng University Preprint (1994).
\bibitem{Neu} M. Neubert, CERN Preprint \#7396 (1994).
\bibitem{Kis} V. V. Kiselev, IHEP Preprint \#94-77 (1994).
\bibitem{Ols} M. G. Olsson and S. Veseli, Madison Physics Preprint \#851.
\bibitem{Gup} A. A. El-Hady, K. S. Gupta, A. J. Sommerer, J. Spence and
J. P. Vary, Iowa State University Preprint (1994).
\bibitem{CSKim} C. S. Kim and A. D. Martin, Proceedings of the International
Workshop on B-Physics, Nagoya, Japan (1994).
\bibitem{Alt} C. Altarelli, N. Cabbibo, G. Corbo, L. Maiani and
 G. Martinelli, {\em Nuclear Physics} {\bf B208} (1982), 365.
\bibitem{Bar} V. D. Barger and R. J. N. Phillips, {\em Collider
Physics}, Addison-Wesley Publishing Company, California (1987).
\bibitem{A1} ARGUS Collaboration, H. Albrecht {et al}, {\em Phys. Lett.}
{\bf B249} (1990), 359-365.
\bibitem{A2} ARGUS Collaboration, H. Albrecht {et al}, {\em Phys. Lett.}
{\bf B318} (1993), 397-404.
\bibitem{C1} CLEO Collaboration, P. Avery {et al} CLEO Conference \#94-7
(1994).
\bibitem{Kim} V. Barger, C. S. Kim and R. J. N. Phillips, Madison
Physics Preprint \#501 (1989)
\bibitem{Suz} M. Suzuki, {\em Nuclear Physics} {\bf B258} (1985), 553.
\bibitem{A3} ARGUS Collaboration, H. Albrecht {et al}, {\em Z. Phys.}
{\bf C57} (1993), 533-540.
\bibitem{who1} ALEPH Collaboration, D. Buskulic {\em et al}, CERN Preprint
\#94-173 (1994).
\bibitem{who2} OPAL Collaboration, R. Akers {\em et al}, CERN Preprint
\#95-02 (1995).

\end{thebibliography}
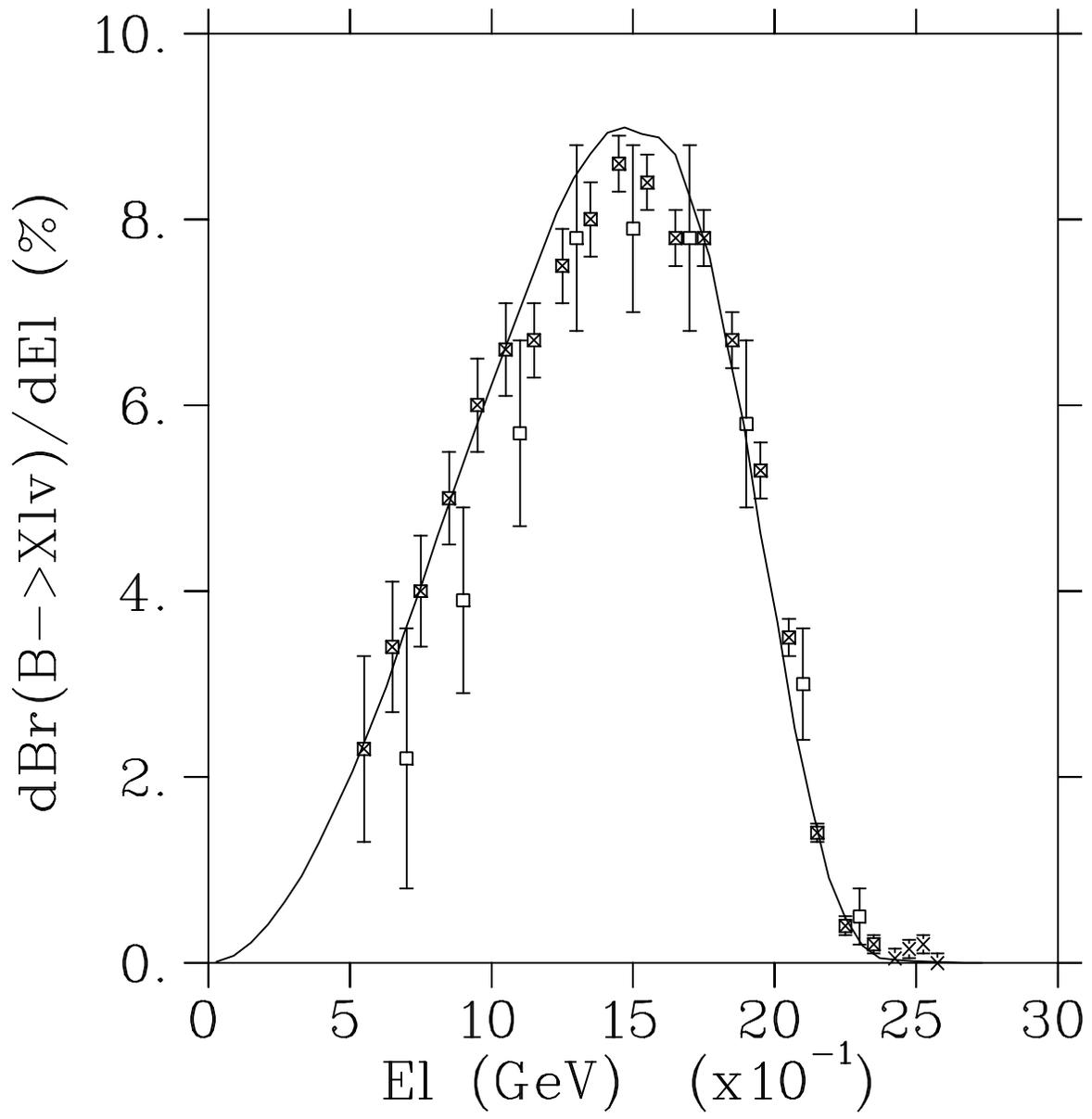
\end{document}